# Topological electronic structure and electronic nematicity in candidate kagome superconductors, $A$Ti$_3$Bi$_5$ ($A$ = Rb, Cs)

Yong Hu[1,*], Congcong Le[2,*], Xianxin Wu[3,*], and Ming Shi[4,*]

[1]Center of Quantum Materials and Devices & Department of Applied Physics, Chongqing University, Chongqing 401331, China
[2]RIKEN Interdisciplinary Theoretical and Mathematical Sciences (iTHEMS), Wako, Saitama 351-0198, Japan
[3]CAS Key Laboratory of Theoretical Physics, Institute of Theoretical Physics, Chinese Academy of Sciences, Beijing 100190, China
[4]Center for Correlated Matter and Department of Physics, Zhejiang University, Hangzhou 310058, China

*To whom correspondence should be addressed:
Y.H. (yong.hu@cqu.edu.cn);
C.L. (congcong.le@riken.jp);
X.W. (xxwu@itp.ac.cn);
M.S. (shi20001231@zju.edu.cn)

**The newly discovered family of titanium-based kagome metals, $A$Ti$_3$Bi$_5$ (where $A$ can be Rb or Cs), has been found to exhibit non-trivial band topology and fascinating electronic instabilities, including electronic nematicity and potential bulk superconductivity. Distinct from their vanadium-based counterparts ($A$V$_3$Sb$_5$), which display a charge density wave (CDW) phase that already breaks rotational symmetry, $A$Ti$_3$Bi$_5$ shows no evidence of CDW, providing a unique platform to study nematicity in its pure form and its interplay with other correlated quantum phenomena, such as superconductivity. In this review, we highlight recent progress in both experimental and theoretical research on $A$Ti$_3$Bi$_5$ and discuss the unresolved questions and challenges in this burgeoning field.**

Owing to the intrinsic geometric frustration of the kagome lattice, transition-metal-based kagome materials represent a rich frontier for the exploring exotic correlated and topological quantum states [1-7]. Among these, the compounds $A$V$_3$Sb$_5$ ($A$ = K, Rb, Cs), featuring a kagome net of vanadium atoms, have emerged as rare examples of kagome superconductors, offering a versatile platform to investigate the interplay between frustrated lattice geometry, band topology, and electronic correlations [8-13]. $A$V$_3$Sb$_5$ has been found to exhibit intriguing similarities to correlated electronic phenomena observed in high-temperature superconductors, such as charge density wave (CDW) [14-19], electronic nematicity [20-23], and possible unconventional superconductivity [Fig. 1g(i)] [24-31].

The fascinating combination of multiple exotic orders and physical phenomena in $A$V$_3$Sb$_5$ has motivated rapid exploration of this family and new materials with similar structures. Dozens of new kagome candidates, isostructural to $A$V$_3$Sb$_5$, has been predicted via high-throughput first-principles calculations [32]. Among them, a family of titanium-based kagome superconductors, $A$Ti$_3$Bi$_5$ (where $A$ can be Rb or Cs, and KTi$_3$Bi$_5$ has not yet been realized), has been successfully synthesized recently [33-35], featuring a two-dimensional (2D) titanium kagome net within the Ti-Bi sheet (Fig. 1a). However, unlike $A$V$_3$Sb$_5$, transport measurements on $A$Ti$_3$Bi$_5$ show no evidence in magnetic susceptibility and electrical resistivity that could be associated with a CDW state [33-35], which is well-established in $A$V$_3$Sb$_5$. Remarkably, $A$Ti$_3$Bi$_5$ exhibits a possible superconducting ground state and an electronic nematicity with rotational symmetry breaking [35-37]. The rotational symmetry breaking signatures of $A$Ti$_3$Bi$_5$ occur solely within the nematic order, in the absence of the concomitant translational-symmetry-broken CDW seen in $A$V$_3$Sb$_5$. Thus, $A$Ti$_3$Bi$_5$ provides a tantalizing platform to study nematicity in its pure form as well as its interplay with orbital degrees of freedom.

In this review, we highlight the recent advancements in studying the novel correlated and topological electronic states in titanium-based kagome superconductors $A$Ti$_3$Bi$_5$. This paper is organized as follows: we first discuss the crystal structure, transport properties, and band topology of $A$Ti$_3$Bi$_5$. Second, we present the unique coexistence of flat bands, type-II Dirac nodal lines and non-trivial topology in the titanium-based kagome materials. Third, we demonstrate the intricate orbital characters and intrinsic interorbital coupling in the multiorbital system, derived from polarization-dependent angle-resolved photoemission spectroscopy (ARPES). Finally, we discuss the understanding of the orbital-selective electronic nematicity and superconducting properties of $A$Ti$_3$Bi$_5$. We conclude by highlighting the open issues and providing a future perspective on this research field.

**Crystal structure and band topology**

$A$Ti$_3$Bi$_5$ crystalizes in a layered hexagonal lattice with the space group P6/$mmm$, sharing the same crystal structure as $A$V$_3$Sb$_5$, but with a kagome net of Ti atoms substituting for the V and Bi replacing Sb. The unit cell consists of alternating layers of Ti-Bi sheets and alkali metal layers (Fig. 1a). Magnetic susceptibility and electrical resistivity measurements indicate an onset superconducting transition ($T_c$) at 4.8 K [33,35] in CsTi$_3$Bi$_5$ and 4.2 K [37] in its sister compound RbTi$_3$Bi$_5$ (Fig. 1b and c). Interestingly, no anomalous CDW-like transition is detected in the normal state of $A$Ti$_3$Bi$_5$. The absence of a CDW is consistent with the scanning tunneling microscopy (STM) measurements (Fig. 1d and e) [35,36]. Optical spectroscopic measurements further confirm the absence of CDW in CsTi$_3$Bi$_5$ [38]. Moreover, the STM measurements reveal a pronounced two-fold symmetric scattering signature, indicating spontaneous six-fold rotational symmetry breaking [35,36], generally attributed to the presence of electronic nematicity (Fig. 1d and e) [20]. Angular-dependent magnetoresistivity (AMR) experiments further support the underlying two-fold rotational symmetry [35].

The electronic structure of $A$Ti$_3$Bi$_5$ has been initially established using density functional theory (DFT) [32,33]. The low energy band structure of RbTi$_3$Bi$_5$ with considering spin-orbit coupling (SOC) is primarily dominated by the 3$d$ orbitals of titanium atoms, revealing multiple bands crossing the Fermi level ($E_F$), as illustrated in Fig. 1f. The characteristic band dispersion of the kagome lattice, including Dirac point (DP) around the $K$ (or $H$) point and flat band, is highlighted in red. Examination of the band structure along the out-of-plane momentum also reveals a type-II Dirac nodal line (DNL), as indicated by the green arrow (Fig. 1f).

Compared to its vanadium-based counterpart, $A$Ti$_3$Bi$_5$ is expected to display a much stronger SOC effect due to the presence of heavier Bi atoms. The substantial SOC results in pronounced gaps at the DP, a negligible gap at the DNL, and at trivial band-crossing points (Fig. 1f). Moreover, the strong SOC can generate intriguing non-trivial topological surface states [33]. Further calculations of topological invariant suggest that the direct gap (highlighted by the blue/orange shaded area in Fig.1f) carries a non-trivial $\mathbb{Z}_2$ topological index, and the band inversion between the Bi-$p_z$ and Ti-$d_{z^2}$ bands around the $A$ point is expected to give rise to topological Dirac surface states (TDSSs) [37]. Additionally, quantum oscillation phase calculations also indicate the presence of nontrivial band topology in $A$Ti$_3$Bi$_5$ [39]. Apart from the absence of a CDW, the coexistence of superconductivity, electronic nematicity, and band topology in $A$Ti$_3$Bi$_5$ [Fig. 1g(ii)] is akin to the novel electronic phenomena found in $A$V$_3$Sb$_5$ [Fig. 1g(i)] [8-13,40].

**Flat bands, Dirac nodal lines and topological surface states**

The theoretically identified flat bands, DNL, and TDSSs in $A$Ti$_3$Bi$_5$ have been directly confirmed by ARPES measurements across all members of this family [37,41-43]. The representative kagome flat bands in the photoemission spectrum taken from CsTi$_3$Bi$_5$ are observed around a binding energy ($E_B$) of 0.25 eV (Fig. 2a), showing overall good agreement with the DFT calculations (Fig. 2b). However, the experimental flat bands appear to extend more widely in $k$-space compared to the calculation [37,41,42], as evidenced by the dispersionless feature around the $\Gamma$ point (Fig. 2a). It is worth noting that in a kagome lattice, a perfectly flat band can only be achieved by considering nearest-neighbor hopping, whereas long-range hopping inevitably introduces finite dispersion. Specifically, the flatness of the kagome flat band depends on several factors, such as the magnitude of long-range intra-orbital hopping and the strength of the inter-orbital hopping. In the case of the multiorbital $A$Ti$_3$Bi$_5$, the DFT calculations may overestimate the long-range intra-orbital or inter-orbital hoppings. This could lead to the experimental flat bands being flatter than those predicted by the calculations.

The type-II DNL expected from the band structure calculations is supported by photon-energy dependent ARPES measurements (Fig. 2c) [37,42]. Consistent with the theoretical bands (Fig. 2d), a series of type-II Dirac-like band crossings are observed from the $\Gamma$ - $M$ plane to the $A$ - $L$ plane, clearly indicating a type-II DNL (marked by the red dashed line). The type-II DNL, attributed to the Ti-$d$ orbitals, is protected by mirror symmetry in the absence of SOC. Upon inclusion of SOC, a tiny gap of approximately 9 meV opens at the type-II DNL due to the small SOC of the Ti atoms. However, this gap is negligible, allowing the DNLs to persist even with SOC.

The TDSSs derived from the bulk non-trivial topology around Brillouin zone center ($\Gamma$) are also identified experimentally [37,41,42]. As shown in Fig. 2e, the ARPES spectra, along the $\overline{\Gamma} - \overline{M}$ direction, reveal Rashba-like bands around the $\Gamma$ point. Photon-energy dependent measurements show that these bands do not disperse with respect to photon energy, distinguishing them from bulk states and indicating their 2D surface nature. The shape of the observed Rashba-like bands and their connection with the bulk bands are consistent with theoretical calculations for the surface states (Fig. 2f), further suggesting the existence of TDSSs with the DP located around $E_B$=0.8 eV (Fig. 2e). However, due to the high sensitivity of the energy position and size of surface states to surface environment details, the energy position of the DP in the experiment is deeper than in the calculations.

**Intricate orbital characters and intrinsic interorbital coupling**

The titanium-based kagome superconductors $A$Ti$_3$Bi$_5$ not only exhibit rich non-trivial band topology but also feature a distinct 3$d$ electronic configuration. Figure 3a presents the ARPES intensity mapping of Fermi surface (FS) taken from RbTi$_3$Bi$_5$. The FS consists of five sheets: one circle-like and two hexagonal-like electron pockets near the $\overline{\Gamma}$ point, one rhombic-like electron pocket at the zone boundary ($\overline{M}$), and one triangle-like electron pockets near the zone corner ($\overline{K}$). The bands crossing $E_F$ are denoted as α, β, γ, and δ, respectively [Fig. 3a, b(i), and c(i)]. Initial orbital-resolved DFT band calculations suggest that the electron-like α band around the $\overline{\Gamma}$ point is contributed by the Bi-$p$ orbital, while the rest of the bands are dominated by the Ti-$d$ orbitals. According to the selection rules in the photoemission process, the band structures can be selectively detected depending on their symmetry with respect to given mirror planes of the experimental geometry. To demonstrate the orbital characters of the multiorbital kagome system, polarization-dependent ARPES measurements have been performed along two representative high-symmetry paths, i.e., $\overline{\Gamma} - \overline{M}$ and $\overline{\Gamma} - \overline{K}$ directions [Fig. 3b(ii, iii) and c(ii, iii)]. A detailed analysis on the polarization-dependent ARPES spectra (Fig. 3a and b) reveals change in orbital characters along the FS contributed by the Ti-$d$ orbitals [37]. Specifically, the β (γ) bands along the $\overline{\Gamma} - \overline{M}$ and $\overline{\Gamma} - \overline{K}$ directions mainly comprise the $d_{yz}(d_{xy})$ and $d_{xz}(d_{x^2-y^2})$, respectively, which agree well with the calculated orbital-resolved band dispersion considering the inter-orbital coupling between the non-degenerate orbitals ($d_{xz}/d_{yz}$ and $d_{xy}/d_{x^2-y^2}$) in the kagome lattice, as illustrated in Fig. 3d.

The variation in orbital characters along the FS suggests a strong intrinsic interorbital coupling in the Ti-based kagome metals [37], which is not observed in the V-based kagome superconductors $A$V$_3$Sb$_5$ [44]. Notably, the inter-orbital coupling observed in RbTi$_3$Bi$_5$ resembles that of iron-based high-temperature superconductors [45], such as FeSe (Fig. 3e), where nematicity is also observed without the presence of CDW. However, due to the $D_{2h}$ site symmetry group, the two sets of $d_{xz}/d_{yz}$ and $d_{xy}/d_{x^2-y^2}$ orbitals in $A$Ti$_3$Bi$_5$ are nondegenerate, whereas the $d_{xz}/d_{yz}$ orbitals are degenerate in iron-based superconductors (tetragonal systems). In this sense, the Ti-based kagome metal is unique and distinct from iron-based superconductors.

**Orbital-selective electronic nematicity**

Figure 4a presents the Fourier transform (FT) of a low-temperature STM topograph, revealing noticeable twofold symmetric electronic signatures [35,36]. This FT intensity anisotropy is more pronounced along specific scattering wave vectors corresponding to scattering between $p_z$ and $d_{xy}/d_{x^2-y^2}$ orbitals (q$_2$) and between $d_{xy}/d_{x^2-y^2}$ orbitals (q$_3$), as labelled in Fig. 1e and Fig. 4b.

Additionally, the experimental FT displays a series of elongated quasi-one-dimensional features near the BZ edge, similar to the quasi-one-dimensional vectors observed in CsV$_3$Sb$_5$ [16,26]. These features and the specific wave vectors contributing to the intensity anisotropy can be explained by the interplay of scattering and interference of electrons primarily from out-of-plane $d_{xz}/d_{yz}$ orbitals and in-plane $d_{xy}/d_{x^2-y^2}$ orbitals (Fig. 4b) [36]. However, Yang et al.'s STM measurements seem to indicate that the quasiparticle interference pattern along $\overline{\Gamma} - \overline{K}$ involving intra $d_{xz}/d_{yz}$ orbitals scattering has stronger six-fold-symmetry-breaking signatures [35], pointing to an orbital-selective electronic nematicity in $A$Ti$_3$Bi$_5$.

The orbital-selective nature is supported by doping-dependent ARPES measurements [37], as evidenced by the orbital-selective doping effect (Fig. 4d-g): Upon surface potassium doping, the two hole-like bands (ζ$_1$ and ζ$_2$) around the $\overline{M}$ point drops by about 60 and 90 meV (Fig. 4g(i), EDC#1), respectively, while the hole-like θ band near the $\overline{K}$ point moves shift by 320 meV (Fig. 4g(ii), EDC#2), and the kagome flat band (η) along the $\overline{\Gamma} - \overline{K}$ direction only moves down by about 40 meV (Fig. 4f(iii), EDC#3). Orbital-resolved DFT band structures unveil a strong *d*–*p* hybridization around $E_F$ in RbTi$_3$Bi$_5$ (Fig. 4h). More intriguingly, *d*–*p* coupling along the $\Gamma - K$ path is stronger than that along the $\Gamma - M$ path, with the θ band around the $K$ point showing the strongest *d*–*p* coupling. This aligns with the anisotropic scattering from $d_{xz}/d_{yz}$ and $d_{xy}/d_{x^2-y^2}$ orbitals (Fig. 1e and Fig. 4a) [35,36], the pronounced energy shift observed in the bands along the $\overline{\Gamma} - \overline{K}$ upon doping [Fig. 4g(ii)] [37], and the rotational symmetry breaking signature associated with specific bands in AC-ARPES measurements [42]. These results suggest an important role of *d*–*p* hybridization in promoting nematicity.

Taking into account the observed interorbital coupling and *d*–*p* hybridization, intra-orbital and inter-orbital bond orders (Fig. 4i-j) have been theoretically proposed to understand the twofold symmetric electronic signature in $A$Ti$_3$Bi$_5$ (Fig. 1e and Fig. 4a-c) [37]. It's suggested that the effective hopping between d orbitals through Bi $p_x/p_y$ orbitals can become nematic once the degeneracy of the $p_x/p_y$ orbitals is lifted. Strong *d*–*p* hybridization can make the Ti-3*d* orbitals more extended, thereby enhancing the non-local Coulomb interaction, which can promote the nematic bond orders. Importantly, these bond orders can break six-fold rotational symmetry while preserving two-fold rotational symmetry, leading to noticeable momentum-dependent nematic features consistent with experimental observations. Furthermore, theoretical work shows that an odd-parity quadrupole order may emerge in CsTi$_3$Bi$_5$, originating from the paramagnon interference mechanism among spin fluctuations of different sublattices, rather than from electron-lattice coupling mechanism [46]. Crucially, the quasiparticle interference signal induced by this odd-parity quadrupole order exhibits drastic nematic anisotropy, consistent with the STM measurements.

**Superconductivity**

Both CsTi$_3$Bi$_5$ and RbTi$_3$Bi$_5$ have been found to exhibit superconductivity at ambient pressure, with the onset of the $T_c$ being higher than that of the vanadium-based kagome superconductors $A$V$_3$Sb$_5$, but with a low upper critical field [35,37]. However, the superconductivity in the titanium-based kagome system has yet to be confirmed by other studies [33,34,36,46,47]. While the ambient-pressure superconductivity remains under debate, recent reports have observed superconducting transition under pressure in CsTi$_3$Bi$_5$ [48]. Interestingly, a double-dome superconductivity is revealed in the temperature-pressure phase diagram of CsTi$_3$Bi$_5$ (Fig. 5a), reminiscent of observations in $A$V$_3$Sb$_5$ (Fig. 5b) [28-30]. Despite the remarkable similarity between the phase diagrams of $A$Ti$_3$Bi$_5$ and $A$V$_3$Sb$_5$ (Fig. 5), the underlying physics of the two superconductivity domes in these kagome superconductors may differ.

In $A$Ti$_3$Bi$_5$, the first superconductivity dome is likely related to some competing order, such as nematic order, while the second dome may be due to variations in electronic structure caused by a dimensional crossover under high pressure [48]. In contrast, in $A$V$_3$Sb$_5$, the SC-I dome is suggested to be strongly linked to CDW instability, while the SC-II dome is attributed to pressure-enhanced electron-phonon coupling [28-30]. Additionally, the Ti atoms have one fewer 3$d$ electron than V atoms, leading to only one van Hove singularity (VHS) near the $E_F$ in $A$Ti$_3$Bi$_5$, which may affect its superconducting properties. However, few experimental measurements have been conducted on the superconducting states of $A$Ti$_3$Bi$_5$. The exception is STM measurements, which reveal a U-shape SC gap, leaving the nature of its superconducting still undetermined [35].

**Summary and perspective**

The titanium-based kagome family, $A$Ti$_3$Bi$_5$, represents as a unique platform to study novel topological states and exotic correlated phases. Despite some similarities, $A$Ti$_3$Bi$_5$ exhibits distinct characteristics in its band structures compared to $A$V$_3$Sb$_5$:

(i) **Strong spin-orbital coupling:** The substitution of Sb atoms with Bi atoms results in a much stronger SOC effect in $A$Ti$_3$Bi$_5$, which may potentially promote richer topological physics in titanium-based kagome superconductors [37,49-51].

(ii) **Absence of lattice instability:** Unlike $A$V$_3$Sb$_5$, which has a pronounced lattice instability contributing to the emergence of the unconventional CDW phase, no such instability has been observed in either experimental or theoretical phonon spectrum for $A$Ti$_3$Bi$_5$ [32,42,49,52].

(iii) **Van Hove singularities:** $A$V$_3$Sb$_5$ hosts multiple VHSs near the $E_F$ [44,53,54], while theoretical calculations on $A$Ti$_3$Bi$_5$, without including SOC, show only one VHS located well above $E_F$.

(iv) **Absence of CDW:** In sharp contrast to $A$V$_3$Sb$_5$, $A$Ti$_3$Bi$_5$ shows no evidence of CDW, likely due to its lack of lattice instability and the VHS being positioned well above the $E_F$.

(v) **Strong $d$-$p$ hybridization:** Due to the larger radius of Bi compared to Sb, the $d$-$p$ hybridization in Ti-based kagome superconductors is anticipated to be significantly stronger [37]. This strong $d$-$p$ hybridization observed in the Ti-based kagome superconductors can extend the Ti-3$d$ orbitals, leading to an enhanced non-local Coulomb interaction, which may promote nematic bond order.

(vi) **Prominent inter-orbital coupling within the kagome lattice:** The Ti-based kagome system exhibits an intricate change in orbital character along the $\overline{\Gamma} - \overline{M}$ and $\overline{\Gamma} - \overline{K}$ directions, suggesting an intrinsic inter-orbital coupling within the Ti-based kagome lattice that is not observed in V-based kagome superconductors [37,44].

In the aspect of electronic nematicity, $A$Ti$_3$Bi$_5$ stands out compared to other families of materials. In most other systems where evidence for electronic nematicity is observed, such as cuprate high-temperature superconductors and iron-based superconductors, the strong electronic correlations and proximity to magnetism make it challenging to determine whether the nematic order is driven by spin, orbital or lattice degrees of freedom. However, the nematicity in $A$Ti$_3$Bi$_5$ occurs in the absence of a CDW order and displays an orbital-selective behavior, pointing towards an electronically driven nematic state. On the other hand, in hexagonal kagome systems, the nematic order has a unique three-state Potts character [55-57], in contrast to the Ising nematicity found in tetragonal lattice. Moreover, due to the $D_{2h}$ site symmetry group, the two sets of $d_{xz}/d_{yz}$ and $d_{xy}/d_{x^2-y^2}$ orbitals in $A$Ti$_3$Bi$_5$ are nondegenerate, whereas the $d_{xz}/d_{yz}$ orbitals are degenerate in tetragonal systems like iron-based superconductors. This makes onsite orbital splitting-induced nematicity irrelevant in multi-orbital kagome metals. Therefore, Ti-based kagome metals offer a fresh opportunity to understand Potts nematicity and its correlation with the electronic orbitals within the realm of hexagonal systems.

The most intriguing issues in $A$Ti$_3$Bi$_5$ warranting future investigations would be its nematicity, superconductivity, and their interplay. To date, nematicity has mainly been identified through STM measurements. The electronic nematicity is expected to couple with lattice degrees of freedom, which allows for its identification via elastoresistance measurements, though such measurements have yet to be conducted. Currently, superconductivity remains under debate, and further solid experimental evidence is necessary for conclusive confirmation. If superconductivity is confirmed at ambient pressure, it would also be of high interest to explore the following in future experiments:

(i) **Superconducting nature and mechanism of superconductivity:** What is the nature of superconductivity, including the superconducting gap and pairing symmetry? What is the underlying mechanism of superconductivity in this family of kagome superconductors?

(ii) **Interplay between electronic nematicity and superconductivity:** Exploring if and how electronic nematicity in the normal state affects Cooper pairing in the superconducting states [36,58]. Will nematic superconductivity be identified?

(iii) **Pair Density wave and:** Will a pair density wave, as observed in $A$V$_3$Sb$_5$ [26], also be found in $A$Ti$_3$Bi$_5$?

(iv) **Topological superconductivity:** Given the strong SOC and identified nontrivial topology in $A$Ti$_3$Bi$_5$, will topological superconductivity be realized in this system?

These questions represent exciting frontiers for future research and could significantly advance the understanding of the interplay between nematicity and superconductivity in kagome-related materials.

**ACKNOWLEDGEMENTS**

The authors would like to thank Haitao Yang, Xiaoli Dong, Jiangping Hu and Andreas P. Schnyder for enlightening discussion. Y.H. acknowledges the support from the start-up fund from Chongqing University (Grant No. 0013045203003) and the Fundamental Research Funds for the Central Universities. X.W. is supported by the National Key R&D Program of China (Grant No. 2023YFA1407300) and the National Natural Science Foundation of China (Grant No. 12047503).

**AUTHOR CONTRIBUTIONS**

Y.H. organized the sections and wrote the manuscript with valuable input from C.L. and X.W.. M.S. oversaw the overall structure of the manuscript. All authors contributed to the discussion and comment on the paper.

**ADDITIONAL INFORMATION**

Competing interests: The authors declare no competing interests.

**DATA AVAILABILITY STATEMENT**

Data availability statement All data needed to evaluate the conclusions in the paper are present in the paper.

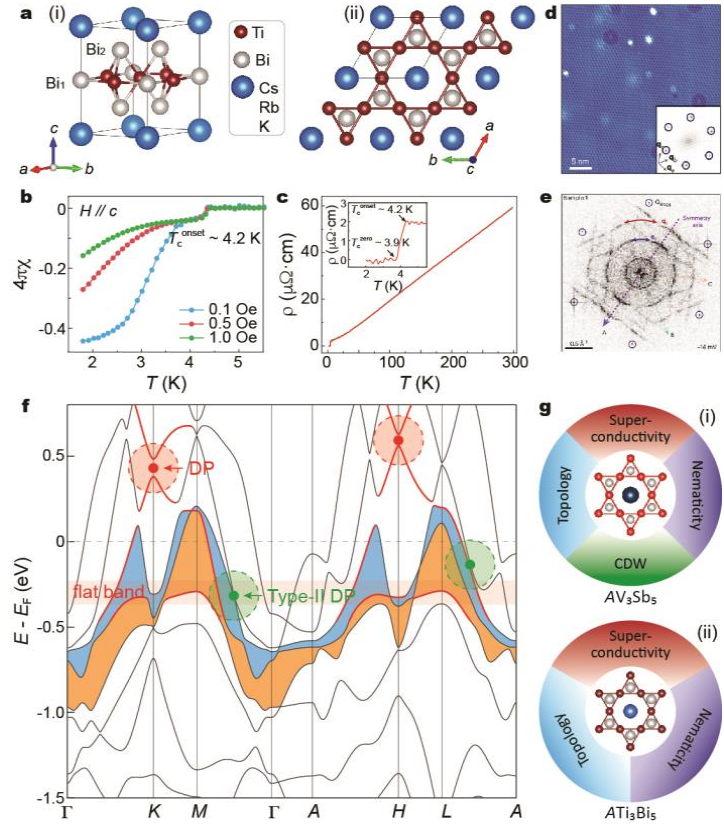

**Fig. 1 | Crystal structure, band topology, and absence of CDW in $A$Ti$_3$Bi$_5$. a** Unit cell of $A$Ti$_3$Bi$_5$ (i) and top view showing the Ti kagome plane (ii). **b,c** Magnetic susceptibilities under various magnetic fields for H ∥ c (**b**) and temperature-dependent resistivity under zero magnetic field (**c**), obtained from RbTi$_3$Bi$_5$. **d,e** STM topograph taken over a Bi-terminated region (**d**) (inset: the corresponding FT showing a hexagonal lattice) and fourier transforms (FT) of a normalized differential conductance L(**r**, V = −78 mV) map taken on the Bi surface (**e**). **f** DFT electronic structure with SOC of RbTi$_3$Bi$_5$, along high-symmetry directions. The shaded areas and arrows indicate the DP (red), and type-II DP (green), and the kagome flat bands. **g** Novel electronic states found in $A$V$_3$Sb$_5$ (i) and $A$Ti$_3$Bi$_5$ (ii). Results in (a-c) are reprinted with permission from ref. 37, data in (d,e) are adapted from ref. 36.

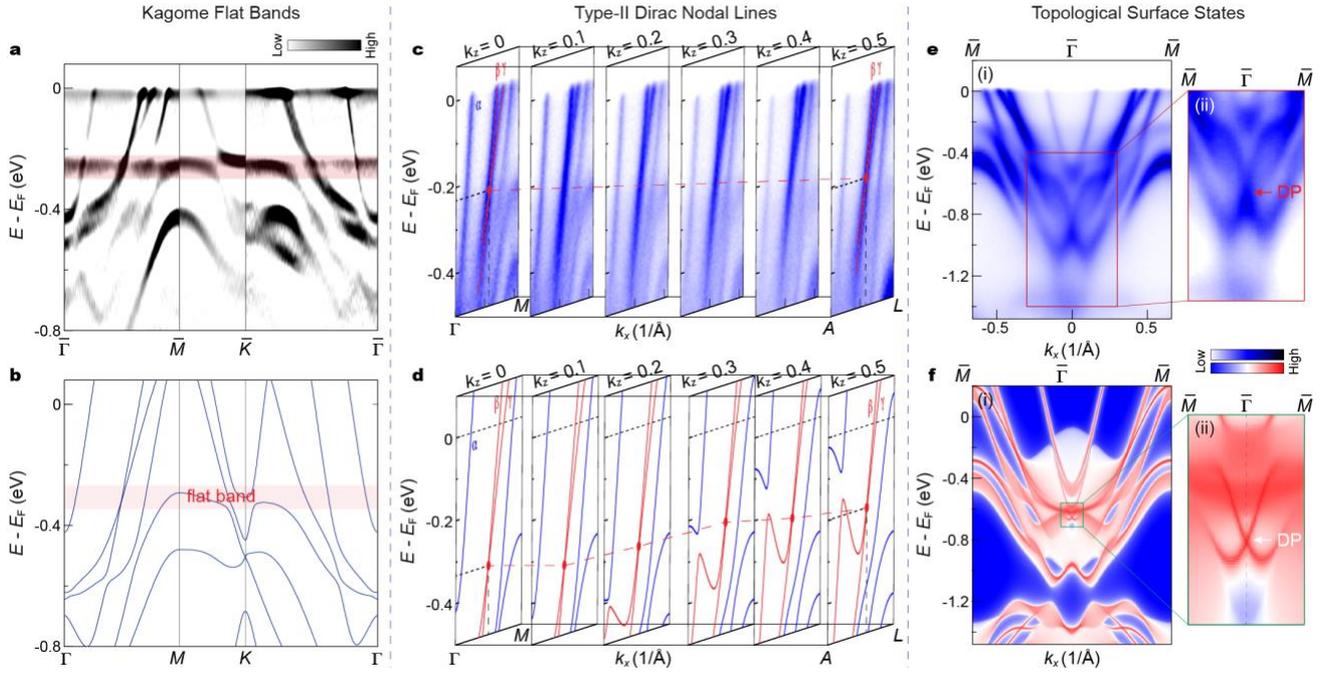

**Fig. 2 | Flat bands, DNLs, and TDSSs in $A$Ti$_3$Bi$_5$. a,b** Experimental (**a**) and DFT (**b**) band structures of CsTi$_3$Bi$_5$ along the $\overline{\Gamma} - \overline{M} - \overline{K} - \overline{\Gamma}$ direction. The red shading highlights the kagome flat bands. **c,d** ARPES intensity plots (c) and calculated band dispersions of RbTi$_3$Bi$_5$, along the $\overline{\Gamma} - \overline{M}$ direction at representative $k_z$ planes (**d**). **e** Experimental band dispersions take from RbTi$_3$Bi$_5$ along the $\overline{\Gamma} - \overline{M}$ direction (i) and a zoomed-in plot of the bands near the DP (ii). **f** Calculated surface spectral function projected on the (001) plane, corresponding to the experimental data shown in (e). Data in (e,f) are adapted from ref. 37.

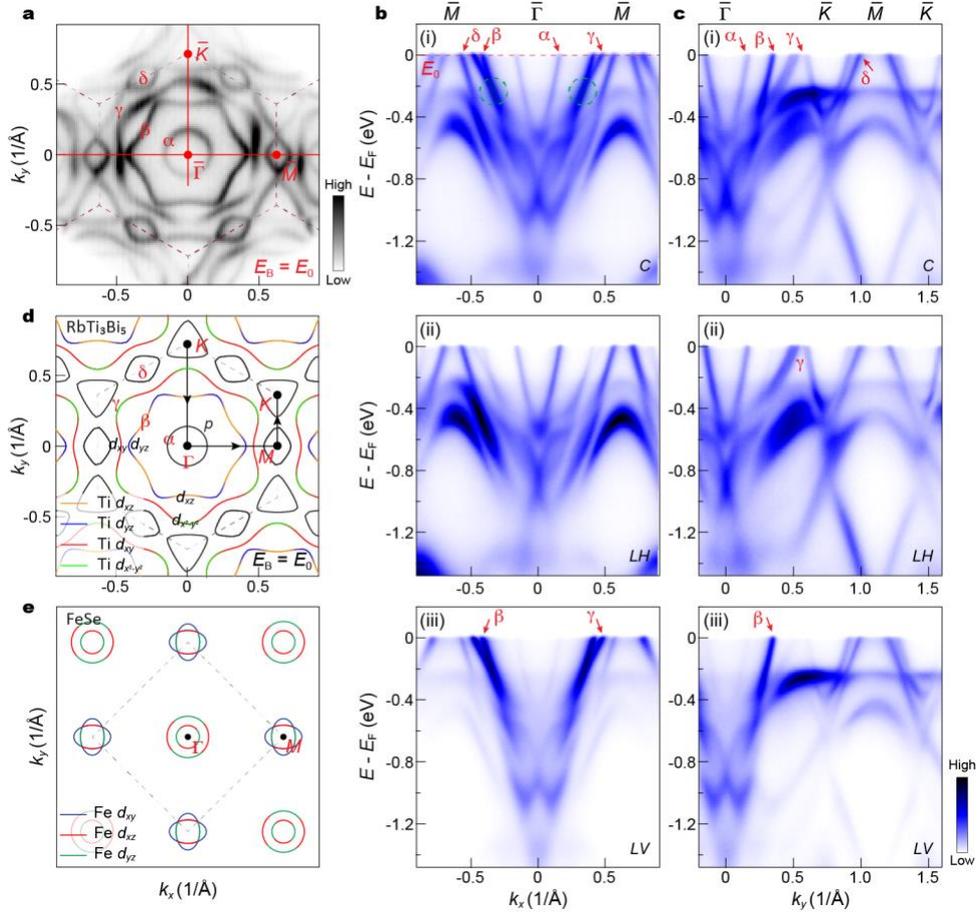

**Fig. 3 | Orbital characters and interorbital coupling in RbTi$_3$Bi$_5$.** **a** Constant-energy contour at the Fermi level. **b** Polarization-dependent ARPES spectra along the $\overline{\Gamma} - \overline{M}$ direction, probed with circularly (C) (i), linear horizontally (LH) (ii), and linear vertically (LV) (ii) polarized light. **c** Same data as in (**b**), but measured along the $\overline{\Gamma} - \overline{M}$ direction. **d** Calculated FS illustrating the change in orbital characters of the Ti-*d* orbitals. Data in (a-d) are reprinted from ref. 37. **e** Schematic FS of bulk FeSe.

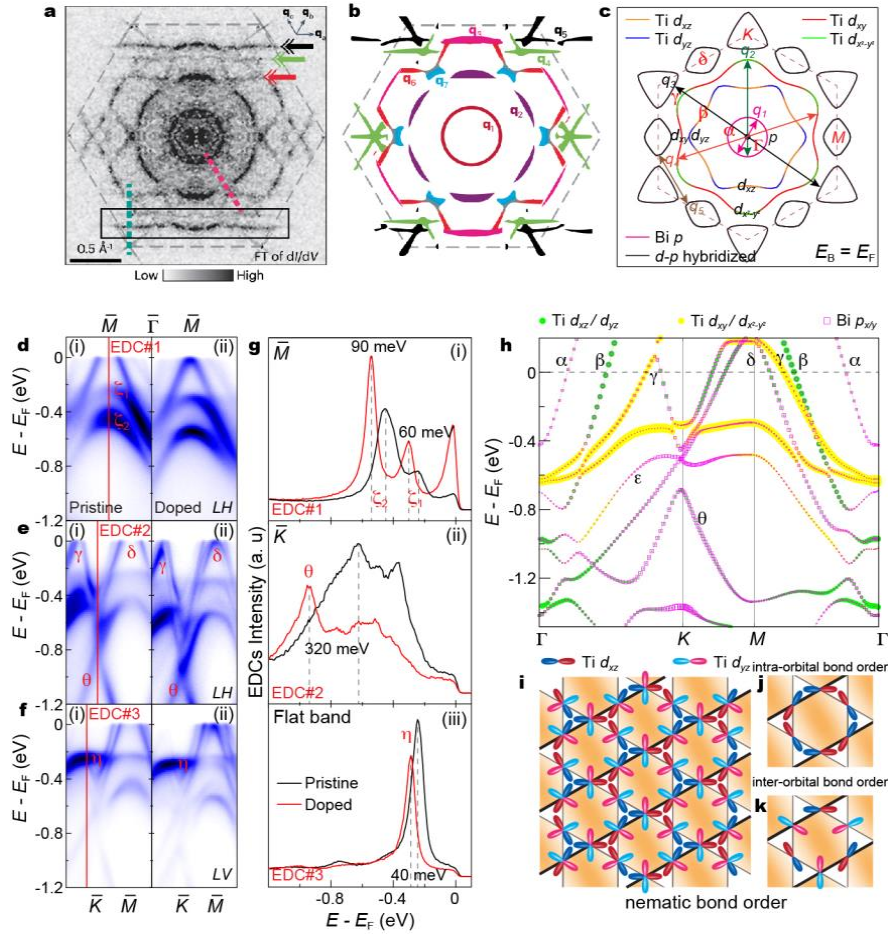

**Fig. 4 | Orbital-selective electronic nematicity and *d-p* hybridization in *A*Ti$_3$Bi$_5$. a** Twofold-symmetrized FT of the normalized differential conductance map for CsTi$_3$Bi$_5$, with the symmetry axis rotated along the *x*-axis. **b** Schematic of main scattering wave vectors in CsTi$_3$Bi$_5$. Results in (a,b) are adapted from ref. 36. **c** Calculated FS displaying four pockets contributed by different orbitals. The scattering vectors ($q_1$–$q_5$) are the wavevectors of the quasiparticle interference patterns in STM measurements [32,33]. **d** Doping evolution of the band dispersions measured on the pristine surface of RbTi$_3$Bi$_5$, along the $\overline{\Gamma} - \overline{M}$ direction, probed with LH polarized light. **e,f** Same data as in (d), but measured along the $\overline{\Gamma} - \overline{K}$ direction with LH (**d**) and LV (**e**) polarizations. **g** Doping-dependent energy distribution curves (EDCs, #1 – #3) of RbTi$_3$Bi$_5$ taken around $\overline{M}$ point (i), $\overline{K}$ point (ii) and flat band (iii), with momentum locations indicated by the red lines in (**d-f**). **h** Bi-*p* and Ti-*d* orbital-resolved DFT band dispersions of RbTi$_3$Bi$_5$. **i-k** Schematic of the inter-orbital and intra-orbital couplings in the Ti-based kagome lattice (**i**), the proposed intra-orbital band order (**j**) and inter-orbital band order (**k**). The yellow background represents the rotational symmetry breaking. Data in (c-k) are reprinted with permission from ref. 37.

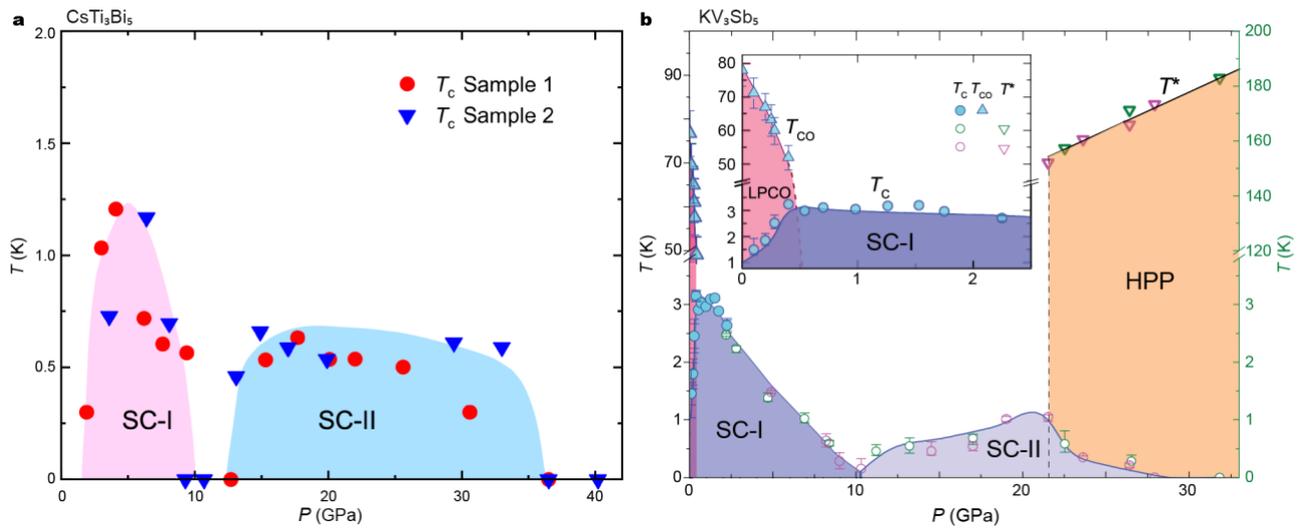

**Fig. 5 | Temperature-pressure phase diagram of $CsTi_3Bi_5$ and $KV_3Sb_5$. a** Two superconducting domes observed in $CsTi_3Bi_5$, including the SC-I phase under lower pressures and the SC-II phase under higher pressures. **b** Two superconducting regions (SC1 and SC2), the low-pressure charge order (LPCO), and the high-pressure phase (HPP) identified in $KV_3Sb_5$. The inset zooms into the low-pressure region, highlighting the possible interplay between superconductivity and the LPCO. Results in (a) and (b) are adapted from ref. 48 and ref. 30, respectively.